\documentclass[psfig]{aa}
\input psfig.tex

\begin{document}                                                                
%-------------------- own definitions -----------------------                   
\def\et{et al.}                                                                 
\def\egs{erg s$^{-1}$}                                                          
\def\egsc{erg s$^{-1}$ cm$^{-2}$}                                               
\def\msu{M$_{\odot}$\ }                                                         
\def\kms{km s$^{-1}$ }                                                          
\def\kmsM{km s$^{-1}$ Mpc$^{-1}$ }                                              
% ------------------------------------------------                              
   \thesaurus{06         % A&A Section 6: Form. struct. and evolut. of stars    
              (03.11.1)}  % Cosmogony,                                         

   \title{An ASCA-ROSAT Study of the Distant, Lensing Cluster A2390}            

   \author{H. B\"ohringer \inst{1}, Y. Tanaka \inst{1, 2}, R.F. Mushotzky\inst{3}
           Y. Ikebe\inst{1}, M. Hattori\inst{4}, } 
                                       
   \offprints{H. B\"ohringer}                                                   
                                                                                
   \institute{$^1$ Max-Planck-Institut f\"ur Extraterrestrische Physik, 
                   D-85748 Garching, Germany\\
              $^2$ Institute of Space and Astronautical Science,
                   Sagamihara, Kanagawa 229, Japan,\\               
              $^3$ Laboratory for High Energy Astrophysics, Code 660,  
                   NASA/Goddard Space Flight Center, Greenbelt, MD 20771, USA\\
              $^4$ Astronomical Institute, Tohoku University,  Aoba Aramaki,
                   Sendai, Japan }

   \date{Received ..... ; accepted .....}                         
   
   \maketitle   
                                                                             
   \markboth {ASCA-ROSAT Study of A2390}{} 

\begin{abstract}                                                             
We present the results of a combined study of ASCA and ROSAT observations
of the distant cluster Abell 2390. For this cluster a gravitational arc
as well as weak lensing shear have been previously discovered.

We determine the surface brightness profile and the gas density distribution
of the cluster from the ROSAT PSPC and HRI data. A combined spatially
resolved spectral analysis of the ASCA and ROSAT data show that the 
temperature distribution of the intracluster medium of A2390 is consistent
with an isothermal temperature distribution in the range 9 to 12 keV
except for the central region. Within a radius of 
$160 h_{50}^{-1}$ kpc the cooling time is found to be shorter than
the Hubble time, implying the presence of a cooling flow. 
In this central region we find strong evidence for a multi-temperature structure. 
Detailed analysis of the combined ASCA and ROSAT data yields a self-consistent 
result for the spectral structure and the surface brightness profile of the
cluster with a cooling flow of about $500 - 700$ M$_{\odot}$ y$^{-1}$ 
and an age of about $10^{10}$ y.

From the constraints on the temperature and density profile of the 
intracluster gas we determine the gravitational mass profile of the cluster
and find a mass of about $2\cdot 10^{15}$ M$_{\odot}$ within a radius of 
$3 h_{50}^{-1}$ Mpc. A comparison of the projected mass profiles of the cluster 
shows an excellent agreement between the mass determined from X-ray data and 
the mass determined from the models for the gravitational arc and the weak 
lensing results. 
This agreement in this object, as compared to other cases where a larger 
lensing mass was implied, may probably be due to the fact that A2390 is more 
relaxed than most other cases for which gravitational lensing mass and X-ray 
mass have been compared so far. 
                                                                                
                   
   \end{abstract}                                                               
                                                                                
%                                                                               
%________________________________________________________________               
                                                                                
\section{Introduction}                                                          
A 2390 is one of the most prominent clusters in the redshift                    
range around z = 0.2.                                                           
It was classified by Abell (1958) and Abell, Corwin \& Olowin (1989)            
only as a richness class 1 cluster. 
As a target of the CNOC survey (Yee \et\ 1996a)       
deep photometric and spectroscopic data were obtained for this cluster,         
and these authors conclude that it should more likely be classified as a           
richness class 3 cluster (Yee \et\ 1996b). This new CNOC data also provide      
a mean redshift for this cluster of z=0.228 (compared to the previous         
literature value of z=0.232 by Le Borgne \et\ 1991).                          
                                                                                     
In X-rays it is among the ten brightest galaxy clusters known at a redshift      
larger than 0.18 (e.g. Ebeling \et\ 1996). It has been observed with            
the EINSTEIN observatory and showed a luminosity of $L_x \sim 1.6 \cdot    
10^{45}$ \egs (in the 0.7 to 3.5 keV energy band) (Ulmer \et\ 1986; if their
result is converted to $H_0 = 50$ km s$^{-1}$ Mpc $^{-1}$ as used 
in this paper) and the cluster has a slightly elongated shape 
(McMillian \et\ 1989).                           
A ``straight arc'' and several arclets were discovered in this clusters         
by Pello \et\ (1991). All these observations underline that A2390 is a very     
rich and massive cluster of galaxies.                                           
                                                                                
Pierre \et\ (1996) have analyzed a deep ROSAT HRI observation and found that      
the X-ray emission from A2390 is very concentrated and highly peaked, 
indicating a strong cooling flow of about 880 \msu y$^{-1}$. The cooling        
flow is centered on the giant elliptical galaxy in the cluster center.          
This together with the observation of the strong lensing features               
may indicate that the very peaked central surface brightness is probably        
the effect of the cooling flow as well as of a steep central                    
gravitational potential in the cluster.                                         
                                                                                
The straight arc has been modeled by Kassiola, Kovner, \& Blandford (1992)      
and Narashima \& Chitre (1993). Pierre \et\ (1996) have also                    
modeled the lensing cluster with an elliptical potential model and             
a second clump in close consistency with the X-ray morphology. They             
find a projected mass within the arc radius of $M(r \le 38'') \sim 0.8          
\cdot 10^{14} h^{-1}$ \msu.  
They compared the lensing mass                      
with the X-ray data by taking the mass profile of the lensing model             
and the gas density profile from the X-ray surface brightness, and calculated 
the temperature profile needed to satisfy the hydrostatic equation. 
The bulk temperatures found by this approach are in the range 
8 to 10 keV. This high temperature is consistent with the large X-ray 
luminosity of the cluster given the generally good correlation between 
X-ray ICM temperature and X-ray luminosity (e.g. Edge \& Stewart 1991).      
                                                                                
A weak lensing shear in A2390 was also observed recently by Squires            
\et (1996). They deduced an elliptical mass distribution, elongated in          
the direction of the straight arc. This is qualitatively consistent        
with the X-ray surface brightness distribution.                                 
As we will show in this paper the mass distribution inferred from               
the X-ray results are in excellent agreement with both the mass deduced from 
the weak lensing analysis and that from the strong lensing modeling.        
                                                                                
The detection of diffuse intracluster light was reported by                     
V\'ilchez-G\'omez \et\ (1994). This may also be taken as a sign that            
the core of the cluster is relaxed and that the debris of tidal                 
stripping of galactic halo material had enough time to settle                   
in the gravitational potential of the cluster.                                  
                                                                                
All these previous studies and the fact that the cluster is very massive and  
X-ray luminous makes A2390 a perfect target for more detailed X-ray 
observations, in particular to compare a more precise mass determination       
from the X-ray data with the optical and lensing results. The indication        
of the fair agreement of the mass in the different previous studies             
and the existence of the strong cooling flow suggest that the cluster is        
essentially relaxed and therefore ideal for the test of the various             
methods of mass determination.                                                  
                                                                                
In this paper we present a combined analysis of deep ASCA and ROSAT PSPC and 
HRI observations of this cluster. The ROSAT observations are discussed in 
Section 2 and the ASCA observations in Section 3. A combined analysis of 
the spectral data of both instruments which provides very interesting evidence 
for multi-temperature structure in the cooling flow region of A2390 
is presented in Section 4. Section 5 contains the results of the cluster mass 
determination from the X-ray data, and compared to the lensing results in 
Section 6.
The cooling flow structure is discussed in detail in Section 7.
Section 8 provides a summary and conclusions. We use a value of 
$H_0 = 50$ km s$^{-1}$ Mpc$^{-1}$ for the Hubble constant throughout this 
paper. Thus 1 arcmin at the distance of A2390 corresponds to a scale of 
277 $h_{50}^{-1}$ kpc.              
                                                                                
%__________________________________________________________________             
                                                                                
\section{ROSAT Observations}                                                    
                                                                                
A2390 was observed with the ROSAT PSPC in November 1993 for 10336 sec.          
About 5100 source photons were registered in the ROSAT hard band (channels      
52 to 201) corresponding to the energy range 0.5 to 2.0 keV. The               
X-ray emission could be traced out significantly to a maximum radius of         
9 arcmin ($2.5~ h_{50}^{-1}$ Mpc). There are a few point sources                
inside the 9 arcmin radius and more point sources surrounding the cluster.      
Taking the brightest five of the inner point sources we find that they          
contribute only about 2\% to the X-ray flux of the cluster in the PSPC hard     
band. We therefore conclude that the contamination by the point sources is 
negligible. The count rate within a 9 arcmin radius (without the 2\% correction) 
of $ 0.49 \pm 0.01$ s$^{-1}$ yields an X-ray flux in the 0.1 to 2.4 energy         
band of $1.2 \cdot 10^{-11}$ \egsc and a cluster rest frame X-ray luminosity  
of $2.7 \cdot 10^{45}$ \egs (0.1 to 2.4 keV) for a hydrogen column density     
of $9\cdot 10^{20}$ cm$^{-2}$ and a temperature of 9 keV (see below and 
Section 3 for the choice of these values). The bolometric luminosity calculated 
with these parameters is about $8\cdot 10^{45}$ erg s$^{-1}$.
                                                                                
An X-ray image of the cluster produced from the PSPC hard band counts is shown 
in Fig. 1. The image has been smoothed by a variable Gaussian filter to show 
the cluster extent and some of the internal structure equally well. The maximum 
filter scale is $\sigma =1$ arcmin. The image has been divided by the exposure 
map and corrected for vignetting and for the background contribution. 
The contour levels are logarithmically spaced. The X-ray surface brightness is 
strongly peaked toward the center, and the cluster image is elongated in the 
SE/NW direction as already observed by Pierre \et\ (1996) in the HRI image. 
For the X-ray maximum of the PSPC image we find the J2000 coordinates, 
21h 53m 37s and 17d 41m 41s, which coincide within 2'' with the position of 
the central cD galaxy and the peak of the X-ray emission in the HRI image 
(Pierre \et\ 1996).  

We also produced an image from the HRI observation for comparison, which        
is shown in Fig. 2. Due to the lower detector sensitivity and higher background, 
the X-ray emission can only be traced to a smaller radius (about 3 - 4 arcmin) 
than in the PSPC image in spite of the larger exposure time of 27.7 ksec. 
The outermost contour in the HRI image corresponds roughly to the sixth contour 
level of the PSPC image. (Note that in Fig. 2, HRI image, the contours are 
linearly spaced). Interestingly, the cluster displays the largest             
degree of ellipticity at an intermediate radius between about 1 and 2.5         
arcmin. Further out the cluster becomes more azimuthally symmetric.             
                                                                                
\begin{figure}                                                                  
\psfig{figure=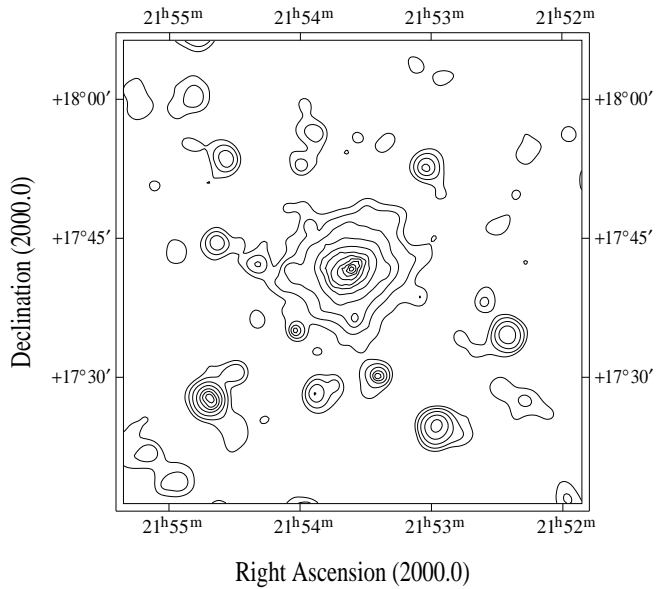,height=8cm}                                         
\caption{
ROSAT PSPC image of A2390 from a pointed observation with an exposure 
time of 10.3 ksec. The image is vignetting and background corrected, divided by 
the exposure map and smoothed with a variable Gaussian filter with a maximum 
width of $\sigma = 1$ arcmin. The contour levels start at a count rate surface 
brightness of 0.16 cts ksec$^{-1}$ arcmin$^{-2}$. The contours are 
logarithmically spaced and increase by factors of 2 (in linear scale).}
                                                                         
\end{figure}                                                                    
                                                                                
\begin{figure}                                                                  
\psfig{figure=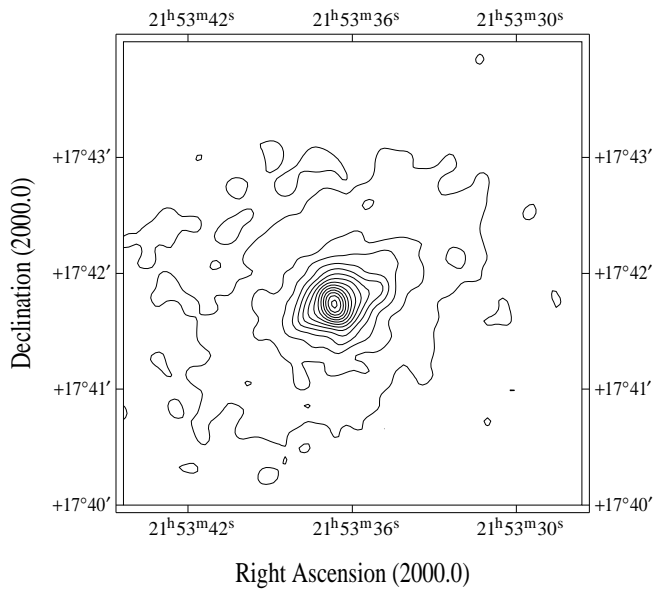,height=8cm}                                          
\caption{Deep ROSAT HRI image of A2390 (see also Pierre \et\ 1996). The image 
has been smoothed with a Gaussian filter with $\sigma = 4$ arcsec.              
For the background corrected image the first contour level corresponds           
to a count rate surface brightness of 9 cts ksec$^{-1}$ arcmin$^{-2}$           
and the levels increase with steps of 13 cts ksec$^{-1}$ arcmin$^{-2}$.}        
                                                                                
\end{figure}                                                                    
                                                                                
The very peaked central surface brightness could either be a sign of            
a very strong cooling flow in the cluster, or it could in principle also      
result from a central point source. To exclude a strong influence               
of a point source we compare the surface brightness profile of the              
cluster in the HRI image with the point spread function (PSF) of the HRI.       
This is shown in Fig. 3, where one notes that the central surface               
brightness peak is not caused by a point source.

\begin{figure}                                                                  
\psfig{figure=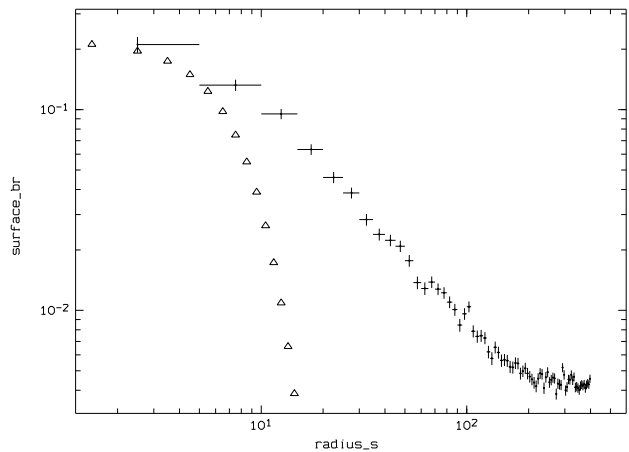,height=6.0cm}                                       
\caption{Surface brightness profile of the central part of the ROSAT            
HRI image of A2390 (crosses). The surface brightness is given in units          
of cts s$^{-1}$ arcmin$^{-2}$ uncorrected for the background. The               
source surface brightness profile is compared to the profile of the point       
spread function of the HRI illustrating that the strongly, centrally            
peaked surface brightness profile is not due to the presence of a                
point source in the cluster.}                                                   
                                                                                
\end{figure}

%_________________________________One column table---------------------------- 
 
   \begin{table} 
   \caption{Results of the $\beta $-model fits to the surface brightnes  
               profiles of the ROSAT PSPC and HRI observations} 

   \label{Tempx} 
      \[ 
         \begin{array}{lllll} 
            \hline
            \noalign{\smallskip} 
         {\rm data}& \ \ S_0   & {\rm core\ radius} & \ \ \ \ \beta & \ \ \ \ \ 
            \chi^2/\nu   \\ 
            & \ \ {\rm cts ~s}^{-1} {\rm arcmin}^{-2} & {\rm arcsec} & &  \\ 
            \noalign{\smallskip} 
            \hline \\
            \noalign{\smallskip} 
   HRI^{a)}  &\ \ 0.18 & 11.5 & \ \ \ \ 0.45 & \ \ \ \ \ \ \ 1.6 \\
   PSPC^{a)} &\ \ 0.18 & 35   & \ \ \ \ 0.62 & \ \ \ \ \ \ \ 1.8 \\ 
   PSPC^{b)} &\ \ 0.23 & 28   & \ \ \ \ 0.60 & \ \ \ \ \ \ \ 2.6 \\ 
            \noalign{\smallskip}
            \hline
         \end{array}  
     \]                                                                        
\begin{list}{}{} 
\item[$^{\rm a}$] uncorrected for the HRI/PSPC PSF  
\item[$^{\rm b}$] corrected for the PSPC PSF  
\end{list}
   \end{table}                                                                  
%                                                                               
%                                                                               
                                                                                
%________________________Table-END______________________________________        

We have determined an azimuthally averaged surface brightness profile           
for the PSPC and HRI observation of A2390.                                      
We fit a $\beta$-model (e.g. Cavaliere \& Fusco-Femiano 1976,                  
Jones \& Forman, 1984) of the form                                              
                                                                                
\begin{equation}                                                                
S(r) = S_0(r) \left( 1 + {r^2 \over r_c^2} \right)^{-3\beta +1/2}               
\end{equation}

\noindent      
to both surface brightness profiles and found that the data can be described 
quite well by these analytical fits out to a radius of about 9 arcmin for 
the PSPC and about 3 arcmin for the HRI. 
The fitting results are summarized in Table 1.    
The PSPC surface brightness profile and the best fit model are shown in Fig. 4. 
We find a best fit core radius of 35 arcsec (close to the angular resolution    
of the PSPC).
A comparison with the HRI profile, also shown in Fig. 4, reveals that the surface 
brightness keeps increasing in the inner part which is unresolved by the PSPC. 
However, the slope of the increase with decreasing radius is different in the inner 
part. This is reflected in the different $\beta $ values found in the PSPC and 
HRI data fits. The good overlap of the HRI and PSPC data at intermediate radii 
shows that the different slopes are not due to an inconsistency in the data.         

In attempting to account for the central surface brightness excess
we have also alternatively performed fits with composite profiles
including a $\beta $-model and a Gaussian peak. We obtain a good fit
with a significantly reduced $\chi ^2$ and $r_c = 47$ arcsec
and $\beta = 0.66 (\pm 0.04)$ which reflects the very slight steepening
of the radial slope at larger radii (see below). This fitted profile
deviates very little from the simple $\beta $-model at $r \ge 30$ arcsec,
confirming the reliability of the simple $\beta $-model fit.

We have also fitted $\beta $-models convolved with the off-axis angle dependent
point spread function (PSF) of the PSPC to the PSPC surface brightness data to
determine the best fitting parameters for the  real surface brightness 
profile. The result of these fits is also given in Table 1. For this
PSF-corrected result a single $\beta $-model is no longer a good fit,  
since the break in the slope seen in the combined HRI/PSPC data is now
also becoming evident in the PSPC profile. If the innermost points are skipped to
determine the best fitting model for radii larger than $\sim $ 30  arcsec, 
we find the same result as obtained from the PSPC data without taking into account 
the PSF.
Therefore, we are using the latter result for the modelling of the cluster
at larger radii. 

Pierre \et\ (1996) obtained the slope parameter $\beta $ of 0.41 in their analysis 
of the HRI data. Their value indicates a slightly flatter central profile than 
our value. The difference is due to the fact that they restricted the fitting 
to the central 250 arcsec radius circle. 
The reason for this striking change of slope will become obvious below. 
We find that this change occurs at around 35 arcsec ($\sim 160$ kpc), 
the cooling radius of the cluster ICM.

Bartelmann \& Steinmetz (1996) claim that the $\beta$-model is not a good 
description of the X-ray surface brightness profile of clusters as found from 
their N-body-hydrodynamic simulations. They argue that the $\beta$-values 
determined in the fit depends critically on the backgound level
and that this introduces a significant bias in the mass determination
built on the $\beta$-model fits. We have carefully checked this problem for the
present data. We find a best fit for the PSPC data of
$ \beta = 0.62$ with a formal error of $\pm 0.022$ (90\% confidence). 
If we perform separate fits to the inner and outer part of the surface
brightness profile we find an indication of a slight steepening of the slope
from $\beta = 0.56$ to $\beta = 0.66$. The corresponding difference in the 
resulting mass profile out to a radius of $3 h_{50}^{-1} $ Mpc is less than 15\% 
for the two extreme limits. Therefore, we can safely conclude that, within such 
an accuracy, the $\beta$-model fit to the surface brightness profile is a 
sufficiently good description for the cluster structure and mass analysis 
presented here.    

We analyze the PSPC spectra in concentric rings around the cluster center      
as defined by the position of the cD galaxy. The temperature for each annulus 
is determined by fiting the Raymond-Smith model. The result is summarized 
in Fig. 5. 
For the fits the element abundances were fixed to 0.3 of the solar 
values (see Section 3). The results clearly show a low temperature in the central 
region of the cluster inside about 2 arcmin ($kT \sim 3$ to 4 keV) as compared to 
the outer region.   
At larger radii the temperature is significantly higher, but due to the         
restricted low energy range of the ROSAT band the temperature cannot be well    
constrained.      
The hydrogen column density, $N_H$, dertermined within 2 arcmin radius is           
$(9.3\pm0.1)\cdot10^{20}$ cm$^{-2}$, which is significantly higher than       
the hydrogen column density derived from 21cm surveys of $ 6.7 \cdot            
10^{20}$ cm$^{-2}$ (Dickey \& Lockman 1990; Stark \et\ 1992).  This discrepancy 
is discussed in Section 7.

\begin{figure}                                                                  
\psfig{figure=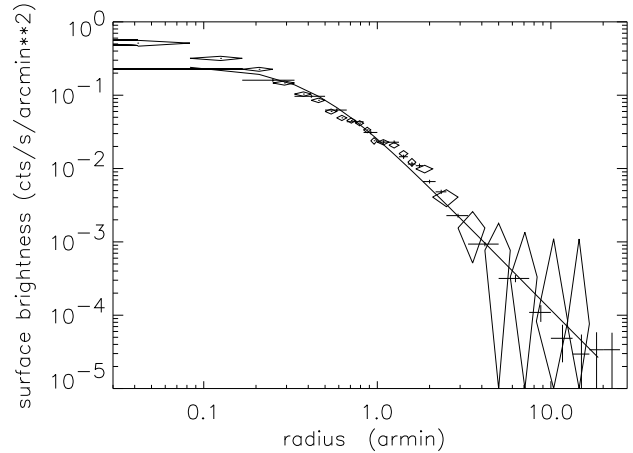,height=6.5cm}                                    
\caption{X-ray surface brightness profile of A2390 determined from the          
ROSAT PSPC observation (crosses). The background is subtracted from the data 
taking into account a combined statistical and systematic error of 10\% for 
the background determination. The fit of a $\beta $-model with the results given 
in Table 1 is shown by a solid line in this plot.                               
The broadening of the profile due to the PSPC PSF is not taken into account in 
this fit. The surface brightness profile deduced from the HRI observation is 
overploted (diamonds) for comparison.    
The HRI surface brightness in cts s$^{-1}$ arcmin $^{-2}$ was multiplied        
by a factor of 2.5 to account for the different detector sensitivity            
for the given spectral temperature and interstellar absorption.}  

\end{figure}

\begin{figure}                                                                  
\psfig{figure=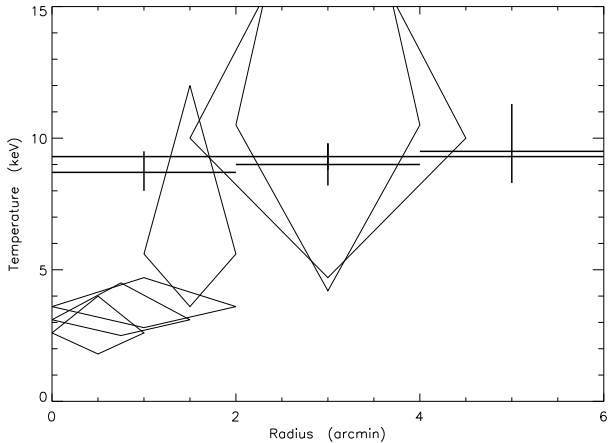,height=6.3cm}  
\caption{Temperatures of the gas in A2390 determined for various radial         
zones by fits of a single-temperature model to the ASCA GIS (crosses) and        
ROSAT PSPC (diamonds) data. The ASCA points are for a fixed $N_H$ at $9\cdot10^{21}$
cm$^{-2}$ (see Table 2). The errors are 90\%-confidence limits.}
                                   
\end{figure}

%---------------------------------------------------------------------          
\section{ASCA Observation}

%_________________________________two column table----------------------------
   \begin{table*}
      \caption{Temperatures determined for concentric rings from the GIS data and 
               from the combined GIS and SIS data}
         \label{Tempx}

     \[
           {\begin{array}{llllll}
            \hline
            \noalign{\smallskip}
{\rm data} & {\rm ring}     & kT    & {\rm abundance}      & N_H    & \chi ^2/\nu   \\
  & {\rm (arcmin)} & {\rm (keV)}& \cdot{\rm solar} & (10^{21} {\rm cm}^{-2}) & \\
            \noalign{\smallskip}
            \hline
            \noalign{\smallskip}
   GIS &    0 - 2 & ~ 8.6 (7.6-9.8)  & 0.26 (0.15-0.36) & 0.95 (0.54-1.4)  & 0.77 \\
   ~~~"  &  ~~~"  & ~ 8.7 (8.0-9.5)  & 0.26 (0.15-0.36) & 0.9~ (fixed)     & 0.76 \\  
   GIS &    2 - 4 & ~ 8.7 (7.7-9.9)  & 0.26 (0.16-0.36) & 1.02 (0.65-1.4)  & 1.04 \\
   ~~~"  &  ~~~"  & ~ 9.0 (8.2-9.8)  & 0.26 (0.15-0.36) & 0.9~ (fixed)     & 1.02 \\  
   GIS &    4 - 6 & ~ 9.7 (7.8-12.4) & 0.31 (0.13-0.50) & 0.84 (0.27-1.5)  & 0.89 \\
   ~~~"  &  ~~~"  & ~ 9.5 (8.3-11.3) & 0.32 (0.14-0.50) & 0.9~ (fixed)     & 0.87 \\
   GIS &    0 - 6 & ~ 9.0 (8.3-9.7)  & 0.27 (0.20-0.34) & 1.04 (0.79-1.3)  & 0.88 \\
   ~~~"  &  ~~~"  & ~ 9.3 (8.8-9.8)  & 0.27 (0.20-0.34) & 0.9~ (fixed)     & 0.88 \\

            \noalign{\smallskip}
            \hline
            \noalign{\smallskip}
   GIS+SIS &  0 - 2 & ~ 9.2 (8.4-10.1) & 0.25 (0.17-0.34) & 0.90 (0.73-1.1) & 0.94 \\
   ~~~" &  ~~~"     & ~ 9.2 (8.6-9.8)  & 0.25 (0.17-0.34) & 0.9~ (fixed)    & 0.94 \\  
   GIS+SIS &  2 - 4 & ~ 9.2 (8.3-10.2) & 0.27 (0.18-0.36) & 1.06 (0.81-1.3) & 1.04 \\
   ~~~" &  ~~~"     & ~ 9.6 (8.9-10.4) & 0.27 (0.18-0.36) & 0.9~ (fixed)    & 1.04 \\  

            \noalign{\smallskip}
            \hline
         \end{array} }  
      \] 
%\begin{list}{}{}
%\item[$^{\rm a}$] This is footnote a
%\end{list}
   \end{table*}

%________________________Table-END__________________

A2390 was observed with ASCA on 20 June, 13 November, and 11 December, 1994,    
for a total exposure time of about 40 ksec. Both the GIS and SIS
data are analysed. We subdivide the region within 6' from the center into 
three regions: the central circle of 2' radius and two annuli with radial ranges 
2' - 4' and 4' - 6' from the center, respectively. The observed spectrum 
for each region is analyzed separately. The SIS data are limited to 4' radius. 
The data from the two GIS detectors, GIS2 and GIS3, and the two SIS detectors, 
SIS0 and SIS1, are respectively added for the analysis. The background data are 
obtained from the accumulated ``blank fields'' data from which the discrete 
sources are removed (Ikebe 1995) for the same radius regions of the detector 
f.o.v. 

A single-temperature Raymond-Smith plasma model gives good fits 
to the data. The results for the GIS data are summarized in Table 2 in which the 
best-fit parameter values and the 90\%-confidence limits (in parentheses) are 
listed. The results for the SIS data in the radius ranges 0' - 2' and 2' - 4' 
are fully consistent with those for the GIS data. The best-fit 
parameters determined from simultaneous fits to both the GIS and SIS data are 
also listed in Table 2. These results show that the plasma temperature is about 
9 keV and the abundances is about 0.3 solar.
The observed GIS and SIS spectra, not corrected for the instrument 
response, of the photons within 2' radius from the cluster center are shown 
in Fig. 6, together with the best-fit model to the combined data set.  
One can clearly see a peak around 5.5 keV which is due to the redshifted iron 
$K_{\alpha}$ lines. 
                                                                                
\begin{figure}                                                                  
\psfig{figure=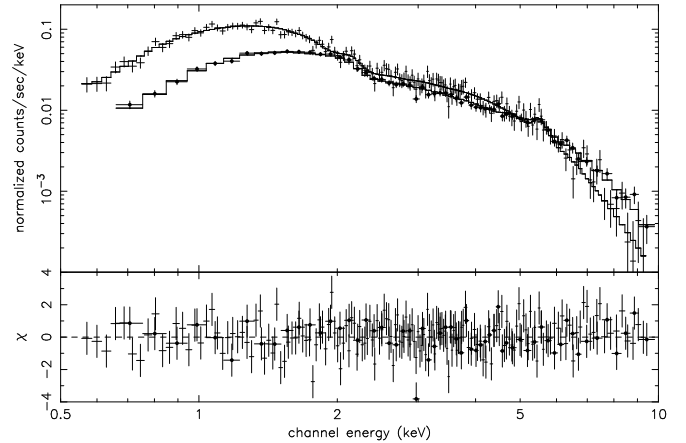,height=7.5cm}                                          
\caption{ASCA GIS ($\bullet$) and SIS (+) spectra inside the central 2'-radius of A2390. 
A single-temperature Raymond-Smith model is fit to both spectra simultaneously. 
The best-fit parameters are given in Table 2. }          
\end{figure}

The total flux within 6' radius is approximately         
$1.4 \cdot 10^{-11}$ \egsc in the range 2 - 10 keV, and the bolometric         
flux is estimated to be $3 \cdot 10^{-11}$ \egsc. Thus obtained bolometric 
luminosity of $8 \cdot 10^{45} h_{50}^{-2}$ \egs is in perfect
agreement with the PSPC result.

%---------------------------------------------------------------------------    
\section{Combined spectral analysis}

The temperature as a function of radius obtained from the fits of a 
single-temperature model separately to the ASCA GIS and the ROSAT PSPC data 
is summarized in Fig. 5. As clearly seen in the figure, 
with use of a single-temperature model, the ROSAT PSPC spectrum          
shows a distinct cooling within 2' of the center. On the other hand, the        
ASCA GIS spectrum within 2' radius shows a much higher temperature than those   
obtained from the PSPC data. This appearant discrepancy suggests that           
the plasma is not of a single temperature but has a more complex temperature    
structure in the inner region. Suppose the plasma comprises components          
of different temperatures. The relative contributions of low-temperature        
components are larger in the ROSAT band ($< 2$ keV) than in the ASCA            
band (1 - 10 keV for the GIS). Hence, the ROSAT PSPC is sensitive to             
the contribution of low-temperature components. However, the ROSAT band         
does not give enough constraint on the high-temperature components. 
Most of the information on high-temperature components is contained in          
the ASCA band. For this reason the simultaneous use of both the ROSAT PSPC data 
and the ASCA GIS data provides maximum constraints on the multi-temperature 
plasma models.

We performed model fitting to the combined PSPC, GIS and SIS data for the 
inner 2' radius. A single set of model parameters is determined from the 
combined data, except for the normalization constants which are determined 
separately for the three instruments because of a possible systematic 
difference between them.                                                              
                                                                                
First, we employ a two-temperature model which assumes that the plasma          
within 2' of the center consists of a high-temperature component at             
$T_H$ and a low-temperature component at $T_L$. The Raymond-Smith model is used 
for each component. The element abundances are assumed to be the same for both 
components.  
A good fit is obtained for a wide range of $kT_L$.            
As an example the fit of the model for $kT_L = 1.0$ keV to the data is shown    
in Fig. 7. The contribution of the low-temperature component in terms            
of the emission measure varies with $kT_L$, from about 4\% for                  
$kT_L = 0.5$ keV to about 15\% for $kT_L = 2$ keV. The best-fit $kT_H$          
increases slightly with $kT_L$, from 9.3 keV                               
for $kT_L = 0.5$ keV to 11.5 keV for $kT_L = 2$ keV. The best fit           
$N_H$ is $(9 \pm 1) \cdot 10^{20}$ hydrogen atoms cm$^{-2}$.  
We note again that this value is somewhat larger than the galactic neutral 
hydrogen column at the cluster position of $6.7 \cdot 10^{20}$ cm$^{-2}$       
(Dickey \& Lockman 1990, Stark \et\ 1992). (See Section 7.)

As shown below, it is not surprising that two-temperature models give a good 
fit to the combined ASCA and ROSAT PSPC spectra. 
To demonstrate that the above results are not produced by a forced fit, the 
two-temperature Raymond-Smith model is applied separately to the ASCA 
(GIS \& SIS combined) data and the ROSAT PSPC data. 
Fig. 8 shows the acceptable ranges (90\% confidence limit) of the temperature 
for the low-temperature component and its normalization factor 
(the emission measure relative to that of the high-temperature 
component), determined respectively for the ASCA and the ROSAT PSPC data. 
In fact, the acceptable regions for the two overlap each other over a wide 
range of $kT_L$. 
Therefore, for the parameter values within the overlap, the model can give 
an acceptable fit to the combined data set.

\begin{figure}                                                                  
\psfig{figure=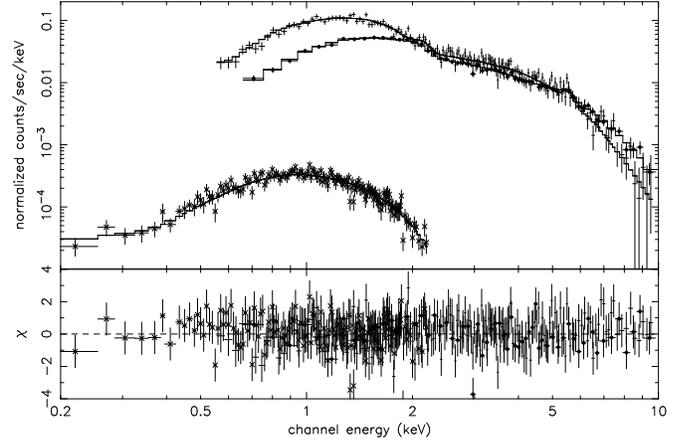,height=7.5cm} 
\caption{Simultaneous fit to the GIS ($\bullet$), SIS (+) and PSPC ($\ast$) 
spectra for the central 2'-radius region of A2390. 
A two-temperature Raymond-Smith model fits to the data with a reduced 
$\chi^2$-value of 0.97. The best-fit parameters are:        
$kT_H = 9.8$ keV, $kT_L = 1.0$ keV (fixed), $Z = 0.26 Z_{\odot}$, 
$N_H = 8.8\cdot 10^{20}$ cm$^{-2}$, $E_L/E_H = 0.05$, where $E_H$, $E_L$ are 
the emission measures of the high- and low-temperature components.}  

\end{figure}

\begin{figure}
\psfig{figure=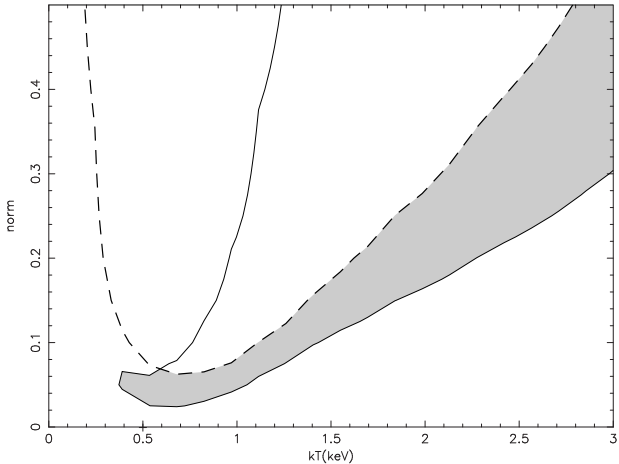,height=6.5cm}
\caption{The 90\%-confidence contours for the two-temperature Raymond-Smith model 
determined separately for the combined GIS and SIS spectra (dashed curve) and for 
the ROSAT PSPC spectrum (solid curve), both within the central 2'-radius of 
the cluster. 
The parameters varied are the temperature of the low-temperature component 
and the normalization factor (the emission measure relative to that of 
the high-temperature component). For the ASCA data $N_H$ was fixed to
$10^{21}$ cm$^{-2}$. For the PSPC data $kT_H$ and the abundances were fixed to 
9.5 keV and 0.3 $Z_{\odot}$, respectively. The acceptable domains are inside the 
solid curve extending to upper right for the PSPC data and below the dashed curve 
for the ASCA data, overlapping in the shaded region.}
\end{figure}

\begin{figure}                                                                  
\psfig{figure=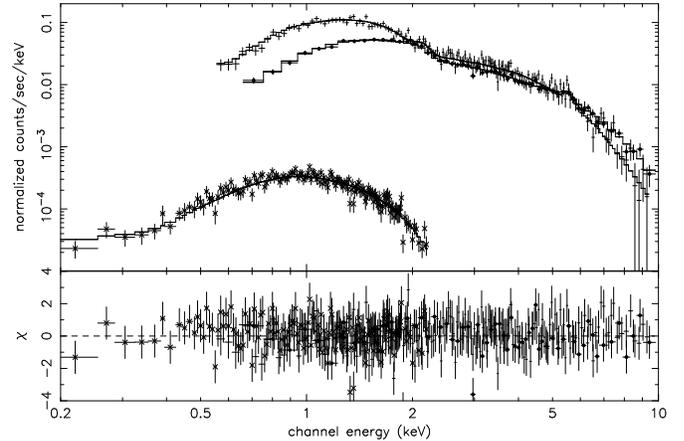,height=7.5cm}
\caption{Simultaneous fit of a cooling flow model (including a contribution of 
non-cooling hot component) to the GIS ($\bullet$), SIS (+) and PSPC ($\ast$) 
spectra for the central 2'-radius circle of A2390. The model gives a good
fit to the combined data (reduced $\chi^2$ of 0.97) with the following 
best-fit parameters: 
$kT_{H} = 12.1$ keV, $\dot{M} \sim 700$ \msu yr$^{-1}$, 
$ Z = 0.3 Z_{\odot} $, $N_H = 8.7 \cdot 10^{20}$ cm$^{-2}$.
}  
\end{figure}

For a more realistic modelling, we employ a cooling flow model developed            
by Mushotzky and Szymkowiak (1988; see also Canizares \et\ 1983; 
Johnstone \et\ 1992), assuming a constant mass flow rate          
throughout the cooling flow and that the differential emission measure          
at a temperature $T$ is proportional to the cooling time              
(hence inversely proportional to the bolometric luminosity at $T$).             
This model includes five parameters: mass accretion rate $\dot{M}$,             
a parameter which describes the distribution of emission measure versus         
temperature, $s$, abundances of the elements, the initial plasma temperature, 
$T_H$, the lowest temperature of the cooling flow relevant for the
X-ray observation, $T_L$. We require pressure equilibrium in the cooling flow, 
which corresponds to the case where $s = 0$. A contribution of emission from 
non-cooling high-temperature (at $T_H$) plasma is also included, considering 
the case in which only a part of the plasma within the projected radius of 2' 
is involved in the cooling flow. In fact, no acceptable fit can be obtained 
without inclusion of a non-cooling component. In the fitting, $T_L$ was fixed 
at 0.5 keV, since the result is practically insensitive to the $T_L$ value         
chosen in the range 0.1 - 1 keV. A good fit is obtained with this model,        
as shown in Fig. 9. The best-fit result shows that $kT_H \cong 12.1$           
keV and $\dot{M} \cong 700$ \msu yr$^{-1}$ with the 90\%-confidence 
range 400--850 \msu yr$^{-1}$ and that still approximately 80\%    
of the observed flux within 2' radius comes from the non-cooling 
high-temperature plasma.

The 2'-radius data alone do not have a statistical accuracy to constrain
$kT_H$ well enough, accepting a fairly wide range up to 18.5 keV
(90\% confidence limit).
A tighter limit on $kT_H$ can be obtained by using the GIS data within the
6'-radius circle that have the best statistical accuracy.
We assume isothermality of the hot component within this radius and include
the central cooling component with $\dot{M}$ determined above.
Employing the telescope response for the brightness profile determined from
the ROSAT data, we obtain the 90\% confidence limits of $kT_H$ to be
$9.5 - 12.6$ keV (with the best-fit  of 11.1 keV) for the acceptable range of
$\dot{M}$. Similar results are also obtained from simultaneous fit of a model
comprising an isothermal hot plasma and a cooling flow to the 2'-radius PSPC
and SIS data and 6'-radius GIS data, thus confirming that the temperature of
the hot component is not much higher than 12 keV.

\section {Mass determination}

From the surface brightness profile of the cluster obtained from the
ROSAT PSPC and HRI observation we can determine the gas distribution
in the cluster. We use the analytical description of the surface brightness
of the $\beta $-model to deproject the surface brightness.
Since the plasma emissivity in the ROSAT band 
in the temperature range of interest (3 to 12 keV)
depends only insignificantly on the
temperature we directly derive the emission measure distribution and the
gas density distribution from these observations. 
The resulting gas mass profile is shown in Fig. 10. 

Under the assumption that the intracluster medium of A2390 is approximately
in hydrostatic equilibrium and that the cluster can be approximated
as being spherically symmetric we can derive the profile of the gravitating 
cluster mass from the derived gas density and temperature profiles
by means of the hydrostatic equation which can be written 
in the following form: 

\begin{equation}
M(r) = -{k T_g(r)~ r \over m_{h}\mu~ G} \left( {d \log T_g(r)\over d \log r} +
{d \log \rho (r) \over d \log r} \right)
\end{equation}

It has been shown by N-body/hydrodynamical simulations of cluster formation
by Schindler (1996) and Evrard \et\ (1996) that the assumption of hydrostatic
equilibrium does in general not lead to a large error ( $\le 15 - 20\% $)
in the mass determination, if cases of strong deviation e.g. such as found  
in cluster mergers are avoided. For A2390 which seems to be a well relaxed 
cooling flow cluster, large deviations are not to be expected. 
It was also demonstrated c.f. in Neumann and B\"ohringer (1996,
see also White et al. 1994, Buote \& Canizares, 1996) that 
ellipticities of the order observed for A2390 can be azimuthally averaged into an 
azimuthally symmetric model without causing an error larger than a few percent
in the derived mass.  
Thus, the main source of error in this analysis is still the 
uncertainty in the temperature profile of the cluster, and the above
applied simplifications are well justified.

To explore the range of possible temperature profiles we have calculated
a series of polytropic models, where the nominal temperature was fixed at
the core radius. The models include the following parameter combinations
which well cover the range allowed by the 
temperature measurements for the models including the cooling flows:
$kT$ = 9.5, 11, and 12.5 keV with $\gamma$ = 0.9, 1.0, 1.1, respectively for
all nominal temperatures. Fig. 10
shows the resulting profiles for the maximal and minimal mass as a function
of the radius. This range of models allows for some temperature variation
such that for example at a radius of 2 arcmin (0.65 $h_{50}^{-1}$ Mpc) temperatures
in the range of 8.8 to 13.5 keV are accepted.
The dotted line gives the mass profile for an isothermal
model with a gas temperature of 11 keV.

As mentioned above, the temperature of the hot component in the inner
1 - 2 arcmin region is not well constrained at the high-temperature side
if a cooling flow model is considered. In the mass models given in Fig. 10,
we considered the temperature of the hot component at the core radius up to
12.5 keV.
This value is about the upper limit of the temperature determined with
the ROSAT PSPC for the next outer ring (1' to 2') which is hardly affected
by the cooling flow region.
In addition, the ASCA GIS data give an upper limit of 12.6 keV within 6'
radius. In fact, due to the broad PSF of the ASCA telescopes, about 70\% of
the photons inside the 6'-radius circle of the observed GIS image come from
the inner 2' of the cluster center (Ikebe 1995). Thus, the temperature of
the hot component within 2' cannot be much higher than 12.5 keV.

At the limiting radius of the X-ray emission, at 2.5 $h_{50}^{-1}$ Mpc, 
we find a cluster mass of $1.9 (1.1 - 3.1) \cdot 10^{15}$ \msu, and extrapolated 
to 3 Mpc the mass increases to $ 2.3 (1.3 - 3.8) \cdot 10^{15}$ \msu. 
For comparison, this cluster has a similar mass to the Coma Cluster whose mass 
is about $1.1 - 2.8 \cdot 10^{15}$ \msu (Briel et al. 1992, see also 
B\"ohringer 1994 for a more comprehensive set of model calculations). 

For the ratio of the X-ray luminous mass to the total mass of A2390
(which sets an interesting lower limit to the baryon mass fraction of 
the cluster) we find for the outer X-ray radius (2.5 $h_{50}^{-1}$ Mpc)
a value 23 (14 - 40) \%. There is a trend of an increasing gas mass fraction
with radius which is due to the fact that the surface brightness
slope parameter $\beta $ is smaller than one. We find the following
gas to total mass ratios for different radii: 9 (7-11) \% at 
175 $h_{50}^{-1}$ Mpc (the arc radius), 18 (13 - 25) \% at 1 $h_{50}^{-1}$ Mpc,
and 24 (14 - 40) \% at 3 $h_{50}^{-1}$ Mpc.  
Note that this indicates the importance of information on the X-ray 
emission out to large radii as obtained by the PSPC observations. 
An extrapolation of the slope of the surface brightness profile found in 
the HRI observation would lead to quite different results: a little smaller 
gravitational mass due to the shallower gas density gradient but a much larger 
gas mass, thus leading to an erroneous gas mass fraction at large radii for 
this cluster.

\begin{figure}
\psfig{figure=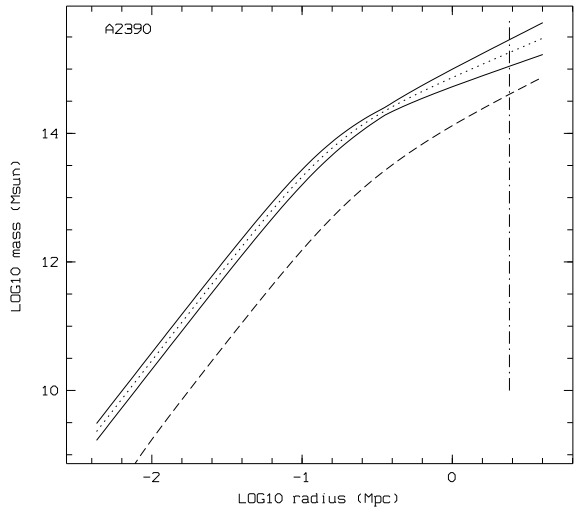,height=7cm}
\caption{Radial mass profile of A2390 determined from the ASCA and ROSAT X-ray
observations. The dashed line gives the gas mass profile determined from the
best-fit $\beta $-model. The solid lines give the upper and lower limits
of the gravitational mass profiles for the cluster as determined from a family
of polytropic models over the temperature range allowed by the X-ray spectral 
analysis. The dotted line in the mass profile corresponds to the isothermal
case for $kT = 11$ keV. The vertical dashed line indicates the radius out
to which X-ray emission is significantly observed, hence any reliable statement 
can be made.}

\end{figure}

%____________________

\section{Comparison to gravitational lensing results}

The mass determined by means of the X-ray observations can now be compared 
to the mass inferred from the gravitational lensing result. 
This is of great interest because both methods are based on completely different 
physical principles and are subject to very different effects. 
For example, the gravitational lensing method does not require the condition 
of virial or hydrostatic equilibrium of the cluster, while it is more sensitive 
to the projection effects of masses in the line of sight. 
Therefore, the comparison of the results obtained from both methods provides
an important confirmation that the assumptions made in the derivation of
the cluster mass are justified.

For the comparison we have to calculate the projected cluster mass profile onto 
the observer plane. We calculate it, cutting the 3-dimensional mass profile at 
an outer radius of 3 $h_{50}^{-1}$ Mpc, which is within the uncertainties equal 
to the virial radius of the cluster. 
The results are shown in Fig. 11 and compared to the lensing results from 
strong lensing by Pierre et al. (1996) and for weak lensing by Squires et al. 
(1996). The agreement is excellent. 
Since the mass models shown in Fig. 10 and 11 are calculated from the 
non-deconvolved PSPC data, we examine the effect of the limited resolution
by calculating a mass model based on the inner HRI surface brightness profile
with an estimated deconvolved core radius of 6 arcsec. The resulting mass profile, 
also shown in Fig. 11, is higher and less steep near the center, but does not 
affect the result at and outside the radius of the arc. 
Similarly, an exact treatment of the multi-temperature structure of the cooling 
flow region would somewhat modify the mass profile. 
However, since the cooling flow also lies inside the arc radius, the mass profile 
will not be affected by the cooling flow at the radii relevant for the comparison 
to the lensing results. 
Therefore, the good agreement is independent of the uncertainties in the mass 
profile due to the limited resolution of the observation and of the cooling flow 
structure. One should note, however, that the weak lensing mass result tends to 
be biased towards a low value, as carefully described in Squires et al. (1996),
due to the fact that not all the cluster mass is covered by the optical observation
on which the lensing analysis is based. Yet, this should lead to a relatively 
small correction, since the mass profile is traced by the lensing result 
already out to a radius of about 1 $h_{50}^{-1}$ Mpc.

\begin{figure}
\psfig{figure=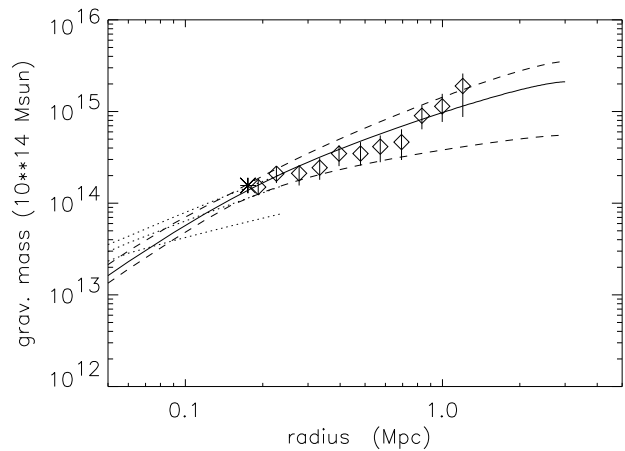,height=6.5cm}
\caption{Radial mass profile of A2390. The mass model is the same as shown in 
Fig. 10, but projected onto the observer plane with a cut-off of three-dimensional 
radius at 3 $h_{50}^{-1}$ Mpc. 
The three dotted lines show the corresponding model calculations when the surface 
brightness fit to the HRI data is used for the inner region. 
The asterisk is the result from the strong lensing model by Pierre et al. (1996), 
and the diamonds show the results of the weak lensing analysis by Squires et al. 
(1996).}

\end{figure}

%_______________________________________________________________________________

\section{Cooling flow}

The first argument for the existence of a cooling flow in A2390 comes
from the calculated short cooling time of less than about $3 \cdot 10^9$ y
for the central region. Adopting a cluster age of $10^{10}$ y, we find
a cooling radius, the radius at which the cooling time equals the cluster age,
of $\sim 160 h_{50}^{-1}$ kpc. We estimate the mass flow rate of the cooling flow
in the standard way, assuming a steady-state, comoving, non-heat-exchanging cooling
flow model, in a simple and more explicit way. In the simple model we 
set the integral radiation loss inside a spherical volume of the cluster equal
to the enthalpy influx

\begin{equation}
\int_0^R n_e n_p \Lambda(T(r))~~ dV(r) = {5kT \over 2\mu m_p} \dot{M}(R)
\end{equation}

where $\Lambda(T)$ is the cooling function, and $\mu m_p$ is the mean particle mass.
One can also calculate the mass flow rate by considering the energy balance in a
differential way also allowing for the energy gain of the gas flowing down the
gravitational potential of the cluster:

\begin{equation}
n_e n_p \Lambda(T) = { \dot{M} \over 4\pi r^2} \left({5k \over 2\mu m_p} 
{dT\over dr} + {d\phi \over dr}\right) + {d \dot{M} \over dr} {1\over 4\pi r^2} 
{5kT \over 2\mu m_p}
\end{equation}

where $\phi(r)$ is the gravitational potential of the cluster. Both approaches 
lead to very similar results. The values for the mass flow from the second formula 
are lower by only about 10-20\% because of the gravitational energy input. 
The estimated mass flow rates at the cooling radius are in the range 550 to 
700 M$_{\odot}$ y$^{-1}$. These results are absolutely consistent with the results 
of Pierre et al. (1996). They get larger numbers for the cooling radius and 
the mass flow rate for an assumed cluster age of $1.4 \cdot 10^{10}$ years. 
The difference between their results and ours is only due to different ages assumed.

As described in Section 4, a mass flow rate around 700 M$_{\odot}$ y$^{-1}$ 
is obtained from the spectral fitting of the multi-temperature cooling flow model 
to the combined ASCA GIS and ROSAT PSPC data. This is in very good agreement with 
the result from the radial surface brightness profile of the X-ray image. 
Thus, the same results have been found in two completely independent ways. 

The cooling flow rate inferred from the imaging data implies in the case
of a steady state a certain residence time of the cooling matter per
intermediate temperature interval. This defines the emission measure
distribution $E(T)$ as a function of temperature, expressed by the equation:

\begin{equation}
E(T) \Lambda (T) dT = {5k \over 2\mu m_{p}} {\dot{M} \over \Lambda(T)}dT
\end{equation}

The normalization of $E(T)$ is the cooling flow parameter determined in the 
spectral analysis.
Therefore, the fact that the two cooling flow rates obtained from the spectral 
and the imaging data agree with each other implies that all the observed properties 
are consistent with the standard model of the cooling flow structure. 
In this case, the cluster must have had a long time ($ \sim 10^{10}$ y) to 
develop this cooling flow.
Otherwise, the steady-state cooling radius should be smaller for shorter times, 
resulting in a smaller cooling flow rate inconsistent with the spectral
result. Similar consistency for such a cooling flow analysis was found
in the study of crystal spectrometer data from the EINSTEIN
Observatory (Canizares et al. 1983) and recently for a similar combined
ASCA-ROSAT study of some clusters by Allen et al. (1996, see also Allen 1997). 

As noted in Section 2, the absorbing hydrogen column density inferred from the 
spectral data (independently from ROSAT and ASCA) indicate a higher value
than measured in 21 cm for the intervening interstellar matter of our galaxy.
To check if this could be absorption caused by cold material in the cooling
flow of the clusters - such as observed in other cluster cooling flows
(e.g. White et al. 1991, Allen et al. 1997, Allen 1997) - we studied the radial
distribution of the excess absorption in detail in the PSPC observation. 
For the radial zones, 0 - 2 arcmin, 2 - 4 arcmin, and 4 - 6 arcmin we
find values for $n_{H}$ of 0.93 (0.84-1.1), 0.81 (0.6-1.2), and an unconstrained 
value in the range (0.3 - 2.0) in units of $10^{21}$ cm$^{-2}$, respectively. 
The range in brackets indicates $2\sigma $ limits. 
The $N_H$-value in the second ring not influenced by the cooling flow is 
still high, but not inconsistent within $2\sigma $ with the galactic value of 
$6.7 \cdot 10^{20}$ cm $^{-2}$ from the 21 cm data (Dickey \& Lockman 1990; 
Stark et al. 1992). This result is consistent with an excess absorption in the 
cooling flow region, though not conclusive.
We note that the absorption values in the Stark et al. (1992) map
were only derived for a grid spacing of two degrees. Therefore, it is still 
possible that the excess absorption found in the X-ray observation of A2390 is 
due to a variation in the galactic hydrogen column. 
An inspection of the IRAS 100$\mu $ maps gives no indication of an enhanced 
interstellar column density in our galaxy in the line of sight of A2390 
with the best estimate for $N_H $ of about $4.6 \cdot 10^{20}$ cm $^{-2}$.

\section{Conclusions}                                                           

The present study of A2390 with combined use of the ROSAT PSPC and ASCA GIS 
data yields important new results.
The two most striking results of the current study are (1) good consistency 
between the cooling flow rates derived independently from the spectral and 
imaging analyses, 
and (2) excellent agreement between the total mass values determined from the X-ray 
data and the gravitational lensing. 
As outlined in the introduction, the cluster shows all the apparant features of 
a well relaxed cluster: 
elliptical symmetry, strong central concentration (and a cD galaxy
which may be the result of this), a cooling flow, and a large intracluster
light halo around the central galaxy. A long cooling time ($\sim 10^{10}$ y) 
implied from the cooling flow rate may be taken as a reinforcement of the picture 
that the cluster was left quite undisturbed for a long time.  

In the morphological analysis of the HRI image of A2390 Pierre et al. (1996)
found some indication of substructure in the cluster. They interpreted
the substructural feature as a trace of substructure in the cluster potential,
and such an excess potential is actually needed in the gravitational lensing
model producing the observed gravitational arc. The subclump has an X-ray
luminosity of only about 1/60 of that of the whole cluster and therefore its
mass is less than about 1/15 of that of the cluster as concluded by Pierre et
al. (1996). Such a small infalling mass component will not cause a
significant disturbance on the equilibrium configuration of the cluster, and also 
may not influence the evolution of the cooling flow seriously. Therefore, the 
presence of this small substructure is not in contradiction to our finding that 
the cluster is generally well settled.

The large measured iron abundance of $\sim 0.3$ of the solar value is in line with 
other observed results for very rich nearby clusters and some distant clusters. 
We should note, however, that lower abundances have been found in 
some rich distant clusters like CL0016+16 (Furuzawa et al. 1997) and 
A851 (Mushotzky \& Loewenstein, 1997, Schindler et al. 1998).

Earlier studies have pointed out cases of striking differences between the lensing 
mass and X-ray determined mass (e.g. Miralda-Escud\'e \& Babul 1995). 
The reason for an excellent agreement between the two in A2390 is most 
probably found in that this cluster is fairly relaxed, as compared to other clusters 
studied by weak lensing technique and X-ray observations. 
For example, A2218 and A2163 show signs of recent merging (Squires et al. 1996b, 
1997). The detailed study of A2218 shows a tendency that the lensing mass is
higher than the X-ray mass, while in A2163 the two mass values are well consistent 
with each other.
PKS0745 which also shows a strong cooling flow and a gravitational arc 
(Allen et al. 1996) may be in a similar situation to A2390, and consistency 
between the lensing mass and X-ray mass could be found in this system, too.
Allen et al. (1996) have already stressed that the agreement of the mass
determination in PKS0745 is most probably the result of the cluster being well
relaxed (see also recent work in Allen 1997).

\begin{acknowledgements}                                                        
H.B. thanks for support by the Verbundforschung under the grant No. 50 OR 93065.
Y.T. is grateful for the hospitality of Max-Planck-Institut f\"{u}r 
Extraterrestrishe Physik where he conducted part of this work. 

\end{acknowledgements}

\end{document}